\documentstyle[11pt,aaspp4]{article}

\def\halpha{H$\alpha$}
\def\stromgren{Str\"omgren}

\slugcomment{To appear in A.J., 2001}

\lefthead{McCullough {\it et al.}}
\righthead{Ionized Filament}

\begin{document}

\title{A Straight and Narrow Ionized Filament}

\author{Peter R. McCullough\altaffilmark{1}}
\affil{Astronomy Department, University of Illinois, Urbana IL, 61801}

\author{Robert A. Benjamin}
\affil{Department of Physics, University of Wisconsin-Madison,
1150 University Ave., Madison, WI, 53706}

\altaffiltext{1}{Cottrell Scholar of Research Corporation.}

\begin{abstract}
We report the discovery of a extremely narrow, extremely linear,
ionized filament. The filament is $\sim 2.5^{\circ}$ long and has an
H$\alpha$ surface brightness of $\sim$0.5 rayleighs.  It is at high
galactic latitude, stretching from
$(l,b)=(140.8^{\circ},39.0^{\circ})$ to $(138.0^{\circ},
37.7^{\circ})$. The filament is approximately ``Y''
shaped. The vertical segment of the ``Y'' is $\sim 1.2^{\circ}$ long
and $\sim 20^{\prime \prime}$ wide, and the widest separation of the
two diagonal segments is $\sim 5^{\prime}$. We discuss four possible
origins for this feature: (1) an extremely low density, nearby jet,
(2) an unusually linear filament associated with some large-scale
nearby nebula, perhaps even the Local Bubble, (3) an ionized trail
left by mechanical input from a star or compact object moving through
the ISM, or (4) an ionized trail left by photoionization (``Fossil
\stromgren~Trail'') from a star or compact object.  We favor this last
hypothesis, and derive some of the basic properties for an ionized
trail. Regardless of whether this latter hypothesis applies to this
specific filament, the basic properties of such a trail, its length,
width, and brightness, are interesting, predictable, and should be
observable behind some white dwarfs.  If the filament is a
photoionized trail, then the source should be closer than a few
hundred parsecs, with a measureable proper motion and a luminosity of
hydrogen ionizing photons of $ ^{<}_{\sim}10^{44} {\rm
ergs~s^{-1}}$. We have searched for such sources in line with the
filament and find one candidate, the X-ray source RXSJ094247.2+700238.
If the M-dwarf binary star GL 360 is the optical counterpart
of the X-ray source, as has been thought, then its proper motion
eliminates it as a candidate, and we have no other potential candidate
to leave a trail.
We note the similarity of this structure to long narrow features
(``canals'') observed as depolarizing regions against the Galactic synchrotron
background and find that this emission filament may also be detectable
as a region of Faraday depolarization.
We suggest future tests for ascertaining the origin
of this filament, and discuss how this structure might be useful to
constrain the thermal and velocity structure of the nearby
interstellar medium.
\end{abstract}


\section{Introduction}

We report the discovery of an unusual object found in the faint,
diffuse, Galactic H$\alpha$ emission: a thin, nearly linear filament
stretching 2.5\arcdeg~long across the sky.  In this paper, we present
our initial discovery and follow-up observations, and explore the
potential explanations for a straight and narrow filament. We favor
the possibility that this filament is produced by the ionizing
radiation from a high proper motion, low luminosity, hot star. However, other
possibilities exist, and we propose specific observations to test this
and alternate hypotheses.

Thuan (1975) developed a theory for the steady-state solution of an OB
star moving through a uniform medium.  Elmergreen (1976) also
addressed the case of an O-type star moving at 100 km~s$^{-1}$ through
a cloudy medium. He determined that a channel of ionized cloud debris
would form around the O star and the channel would be thinnest
perpendicular to the star's direction of motion.  In both cases,
because Thuan and Elmergreen assumed the ionizing source was an O or B
type star, the elongation of the resulting H~II region was not very
large. In order to leave a trail with a large ratio of length to
width, the ionizing source must move a distance much larger than its
\stromgren~radius in a recombination time.  Dupree \& Raymond (1983)
identified white dwarfs as just such a source and coined the term
``Fossil \stromgren~Trail'' (hereafter, FST) to describe the ionized
trail left long after the source had moved far away.  Dupree \&
Raymond predicted that because many white dwarfs are nearby, with
significant proper motions, their FST's might be 20\arcdeg~long. They
also noted because the cooling time for white dwarfs is similar to the
gas' recombination time, an FST could be associated with a (now) cool
white dwarf, one that was recently hot enough to have ionized the FST.
No such FST has been observed in H$\alpha$.

\section{Observations} \label{obs}

On two nights in January 1997, using a 200-mm f/1.8 lens and a 1.8-nm
wide H$\alpha$ filter, we obtained H$\alpha$ images of the sky
containing the galaxies M82 and M81 in order to search for diffuse
H$\alpha$ emission associated with those galaxies.  While that search
turned up negative, we did find a 2.5{\arcdeg}-long, very narrow,
faint, linear feature (Figures 1 and 2). Follow-up observations were obtained
in April and May 1999 with the Mt. Laguna Observatory 1-m telescope
and a different filter that confirmed that the object is not an
artifact of the 200-mm lens or filter (Figure 3).

The filament is a 2.5\arcdeg~long, very linear feature with a
H$\alpha$ surface brightness above background
of $\sim$0.5 rayleighs. (One
rayleigh (1 R) is $4 \pi/10^{6} {\rm photons~s^{-1}~cm^{-2}~ sr^{-1}}$ and
at H$\alpha$~corresponds to an emission measure (EM) of 
approximately 2 cm$^{-6}$ pc.)
The brightness scale was calibrated by comparing our
observations of the Rosette nebula to those of Celnik (1983).  Figures
1, 2, and especially 3 show
that the filament flares at its northern end, forming a ``Y''
shape.  At the southern end of the ``Y'', at declination 71\arcdeg,
the filament is 20\arcsec~wide and at its northern end, at declination
73\arcdeg, it breaks into three filaments. (So the object is not
exactly like a ``Y''; in addition to the two diagonal line segments of
a ``Y,'' it also has a vertical line segment extending between those
two line segments.)  The full width of the ``Y'' at its northern end,
is approximately 5\arcmin.  The length-to-width ratio is thus very
large (2.5\arcdeg/20\arcsec = 450; 2.5\arcdeg/5\arcmin = 30).

We have also detected the filament using the Wisconsin H$\alpha$
Mapper (WHAM) in an imaging mode (Tufte 1997), but the image quality
is poorer than that of Figure 1.  Nevertheless, the WHAM imaging-mode
data give a preliminary velocity of the filament of between
$v_{LSR}=-25~{\rm km~s^{-1}}$ and $0~{\rm km~s^{-1}}$, and confirm
that it is not an extragalactic source.

\section{Possible Explanations} \label{explain}

Using Skyview\footnote{http://skyview.gsfc.nasa.gov/skyview.html,
(\cite{skyview})}, we searched for associated emission in other band
passes. No corresponding structure was seen in the ROSAT All-Sky
Survey (\cite{sno97}), IRAS survey (\cite{whe94}), or 408 MHz
synchrotron emission maps (\cite{has82}).  The 408 MHz maps have much
poorer angular resolution than the H$\alpha$~image in Figure 1, so a
lack of an association may not be compelling in that case.  Sandage
(1976) exposed a red-sensitive Schmidt plate for 2.5 hours to reach a
surface brightness of 27 mag per square arcsecond. Although the
surface brightness of the filament's H$\alpha$~emission alone is too
faint to have been detected on Sandage's plate, the plate shows
reflection nebulosity from interstellar dust (cirrus) quite plainly
throughout the region, and with finer angular resolution than IRAS
data. However, no evidence of the filament is visible on Sandage's
plate.

What is the origin of this object? We consider four possibilities, in
what we consider increasing order of likelihood. The choices are (1)
an extremely low density, nearby jet; (2) an unusually linear filament
associated with some large-scale nearby nebula, perhaps even the Local
Bubble; (3) an ionized trail left by mechanical input from a star or
compact object moving through the ISM; (4) an ionized trail left by
photoionization from a stellar or compact ionizing source (an FST).

\subsection{A low-density, nearby jet?} \label{jet}

Stellar and extragalactic jets can be as straight and linear as the
feature that we have detected (Hughes 1991). However, we consider it
unlikely that the structure we have detected is a jet for three
reasons.  First, as we noted earlier, the filament is a Galactic
source, but we see no clear association with any young star or
molecular cloud.  Second, we see no evidence for associated
synchrotron emission in 408 MHz maps (\cite{has82}), although the
narrow jet would be significantly beam-diluted in one dimension in the
0{\arcdeg}.85~resolution 408 MHz maps.  Third, the H$\alpha$ surface
brightnesses typical of stellar jets exceeds that observed for our
filament by ${\sim}3$ orders of magnitude, e.g.,  the jet associated
with HH 111 has an \halpha~surface brightness of 500 rayleighs
(Reipurth {\it et al.} 1997).
The latter two reasons do not rule out the possibility that the object
is a stellar jet, but if it is one, then it is much fainter than any other
previously observed.

\subsection{Filamentary nebula, straight-and-narrow by chance?} \label{filament}

It is possible that the object is just a natural interstellar filament
and has no stellar origin. Other large scale H$\alpha$ filaments are
seen (Haffner, Reynolds, \& Tufte 1998).  However, our filament is
much straighter than nearly any other nebular structure.  The
``Southern Thread'' near the Galactic center is as narrow and nearly
as straight as our filament, but the ``Southern Thread'' is a
nonthermal source with no detectable recombination radiation and
therefore is a much different physical situation than our filament
(Lang 1999).  Our filament is at moderate galactic latitude, $b
\approx 39^{\circ}$, and is not associated with a traditional HII
region.  Although WHAM has recently enabled the discovery of several large-scale
high-latitude HII regions, we have inspected these maps (Haffner 2000)
and find no clear association between any well-defined nebula and our
filament. Another possibility is that it is associated with the Local
Bubble. However, the fact that we have not seen any other similar
features in any other part of the sky makes this possibility seem
unlikely.

\subsection{Shock-ionized Trail?} \label{shock}

Several examples of interactions between moving stars and interstellar
matter are now known, including bow shocks around O stars and neutron
stars.  The resulting emission structures have been assumed to arise
in the zone of impact between the stellar or pulsar wind with the
ISM. The O-type stars' bow shocks tend to be shaped like arcs or
comets and are not long, narrow filaments (Van Buren et al. 1990).
Because of their large speed, pulsars can leave long, narrow
trails. The clearest example of such is the ``Guitar nebula'', a
bow-shock and trail left behind the fast pulsar PSR 2224+65 (Cordes,
Romani, \& Lundgren 1993).  However, there are no O or B stars in the
vicinity of our object, nor are there any known radio pulsars. This
does not preclude the possibility that a neutron star is responsible,
since we may not be in the pulsar beam path. However, our filament is
$\sim 60$ times fainter than the ``Guitar nebula'', and $\sim 150$
times longer (2.5\arcdeg~ vs.  1\arcmin). For the Guitar Nebula to
appear 2.5\arcdeg~long, it would have to be 13 pc distant, i.e. 4
times closer than the nearest neutron star, RXJ1856.5-3754, which is
60 pc distant.  RXJ1856.5-3754 has an associated H$\alpha$ bow shock
(van Kerkwijk \& Kulkarni 2000); however, it is only $\sim 5^{\prime
\prime}$ wide and $\sim 10^{\prime \prime}$ long, which is a much
smaller size and different morphology than our filament.
  
\subsection{Fossil \stromgren~Trail?} \label{photo}

Finally, we consider that our object might be an example of a ``Fossil
\stromgren~Trail'' (FST) left behind an ionizing source cruising
through the local interstellar medium, as predicted by Dupree \&
Raymond (1983).
Dupree \& Raymond focused on the FST left behind a white dwarf, in an
attempt to explain the detection of interstellar C IV absorption in
white dwarf stars. In principle, any low luminosity ionizing
source could leave such a trail; in this section we use the
abreviation ``WD'' to refer to ``white dwarf'' but also, more
generically, to any source of sufficiently low ionizing luminosity
that its \stromgren~radius is small compared to the distance it
travels in a recombination time.  No such FST has been observed in
H$\alpha$.

We present here some fundamental considerations concerning the width,
length and brightness of an FST.  The principal parameters are the
WD's space velocity, $v~{\rm (km~s^{-1})}$, and its rate of production
of H ionizing photons, $Q_{44}=Q_{H}/10^{44} ~{\rm s^{-1}}$. (We
choose the normalization to match the ionizing luminosity of a hot
(young) white dwarf (Dupree \& Raymond 1983)).  We define a factor to
account for geometrical projection,
$p=~v~/~v_t~=\sqrt{1+v_{r}^{2}/v_{t}^{2}}$, where $v_{r}$ is the WD's
radial velocity, and $v_{t}$ is the transverse velocity. The
transverse velocity can be obtained using the distance,
$d_{100}=d/100~{\rm pc}$, and the proper motion, $\mu ~{\rm
(milliarcseconds~yr^{-1})}$, of the WD: $v_{t}~{\rm (km~s^{-1})}=0.48
\mu d_{100}$. Other parameters include the gas density, $n ~{\rm
(cm^{-3})}$, the electron density, $n_{e}$, and the gas temperature,
$T_{4}=T/10^{4}~{\rm K}$. For a recombination coefficient, $\alpha
\cong 2.6 \times 10^{-13} T_{4}^{-0.8} ~{\rm (cm^{-3}~s^{-1})}$
(Martin 1988), one can estimate a recombination time, $t_{rec} \cong
(n_{e} \alpha)^{-1}$.  

For a stationary WD, the \stromgren~ radius is $R_{s}={\rm (1.5~pc)}
~n_e^{-2/3}Q_{44}^{1/3}$ with a central intensity of $I_{H \alpha}={\rm
(1.1 R)}~ n_e^{4/3}~Q_{44}^{1/3}$. For a WD moving transverse to the line of
sight, the parameters of the FST are as follows. The FST's length is
determined by the recombination time of the ionized gas and the WD's
velocity, $\ell=vt_{rec}$.  The FST's width may be obtained by
assuming that the \stromgren~volume, in which the number of
ionizations balances the number of recombinations, is a cylinder with
a length given above. In that case the width (or diameter of the
\stromgren ~cylinder) is $w=\sqrt{4Q_{H}/n \pi v}$, where $Q_{H}$ is
the hydrogen ionizing luminosity (photons ${\rm s^{-1}}$), and we will
express velocity in units $v_{100}=v/100~{\rm km~s^{-1}}$. The
H$\alpha$ intensity is $I_{H \alpha}=0.363 T_{4}^{-0.9} w n_{e}^{2}
{\rm (rayleighs)}$, and the width is given in parsecs.

Putting the above equations into dimensional units, and including the
projection factor $p$, the FST's angular length is

\begin{equation}
\theta_{obs}=(7.1^{\circ}) T_{4}^{0.7}n^{-1} [p^{-1}v_{100}d_{100}^{-1}],
\end{equation}

where the factor in brackets is proportional to the proper motion.
The FST's angular width is

\begin{equation}
\gamma_{obs} \cong (4.0^{\prime})
Q_{44}^{1/2}n^{-1/2}v_{100}^{-1/2}d_{100}^{-1},
\end{equation}

and its H$\alpha$~surface brightness is 

\begin{equation}
I_{H \alpha} \cong (0.42 R) p
Q_{44}^{1/2}T_{4}^{-0.9}n_{e}(\theta)^{3/2}v_{100}^{-1/2}.
\end{equation}

These formulae do not take into account dynamical evolution of the ionized region; this is discussed further in Section 4. 

If we define $R_{obs}=\theta_{obs}/\gamma_{obs}$, then we can combine equations
1 and 2 to yield an equation that relates the luminosity and velocity
of the ionizing source,

\begin{equation}
Q_{44}=\left(\frac{106.5}{R_{obs}}\right)^{2} p^{-2}n^{-1}T_{4}^{1.4}
v_{100}^{3}.  
\end{equation}
 
Note that the latter equation is independent of distance, but
if we define $J_{obs} ({\rm
rayleighs~arcmin^{-1}})=I(H\alpha)_{obs}/\gamma_{obs}$, then we can combine
equations 2 and 3 to estimate the distance of the source,

\begin{equation}
d_{100}=9.52 J_{obs} p^{-1}n^{-2}T_{4}^{0.9}
\end{equation} 

In this system of equations, we have three observables and six
unknowns.  The observable quantities are angular length $\theta$,
width $\gamma$ and intensity $I_{H \alpha}$. The six unknowns are the
projection factor $p$, the gas density $n$ and temperature $T$, the
source distance $d$, ionizing luminosity $Q$, and transverse velocity
$v_t$.  We may assume that for the FST to be easily detectable, $p
\cong 1$, for solar metallicity photoionized nebulae, $T_{4} \cong
1.0$, and a typical density of diffuse neutral interstellar matter is
1 ${\rm cm^{-3}}$. We thus can constrain the combination of distance,
velocity and luminosity of the source using these expressions with our
observational results. Assuming a fixed length, $\theta_{obs}=2.5$,
$\gamma_{obs}$ ranges from $5^{\prime}$ to $0.3^{\prime}$, which
yields a luminosity in the range $Q_{44}=0.05-12.6
p^{-2}n^{-1}T_{4}^{1.4}v_{100}^{3}$ and distance in the range
$d_{100}=14.3-0.9 p^{-1}n^{-2}T^{0.9}$.

For reasonable parameters, therefore, we expect that the source of the
FST should be rather close, i.e. approximately 100 parsecs. If the
source can be identified and studied, we can use the above equations
(or a more complicated model based on these considerations) to better
constrain the properties of the local interstellar medium. 

\subsection{A Search for the Ionizing Source}

We have used the SIMBAD data base to search for associated point
sources along the filament, in particular white dwarf stars, EUVE
sources, ROSAT point sources, pulsars, and high proper motion stars.
We have found only one potential ionizing source, noted both as a
ROSAT bright point source, RX J094247.2+700238, and as a EUVE source,
EUVE J0942+700 (Chistian et al 1999).  The ROSAT source is located at
${\rm (RA, Dec)}=(145.697^{\circ}, 70.044^{\circ})$, with a position
uncertainty of $14^{\prime \prime}$ (1$\sigma$ radius including
systematic error). The EUVE source position has a much larger
uncertainty $(\sim 150^{\prime \prime})$, but is consistent with the
X-ray source. This source is extremely close to the line traced out by
the filament and is 1\arcdeg~south of the Y-shaped section that we have
observed.

Two papers have identified this ionizing source as one of a pair of
common proper motion M dwarfs: LHS 2176/LHS 2178 (also, GL 360/362)
(Hunsch et al 1999; Marino, Micela, \& Peres 2000). However, LHS 2178
is $33^{\prime \prime}$ (2.4 $\sigma$) away from the ROSAT point
source position, while LHS 2176 is $68^{\prime \prime}$ (4.9 $\sigma$)
away. This is a marginal association. If the M dwarf is in fact tht
ionizing source, there are two lines of evidence that suggest that it
is not resposible for the filament we have observed.

First,  if one of the two M-dwarfs were the ionizing source, we can
use the parallax, brightness, and a model of spectral emission to
estimate the ionizing output, and compare this to the expectations
given above.  Measured (Hipparcos) parallaxes place these M-dwarfs 11.8 and 11.5
parsecs distant, respectively (Perryman et al 1997). The measured
hardness ratio of the X-ray source, assuming a Raymond-Smith model of
a thermal plasma, (Raymond \& Smith 1977) may be used to estimate the
temperature of the emitting plasma. Using the conversion factors in
Figure 1 of Mariono et al. (2000), we find that the observed hardness
ratio corresponds to a plasma temperature of $T=2.8 \times 10^{6}$
assuming no intervening absorption. Using the distance to LHS 2178,
the observed X-ray count rate, and this plasma temperature, we have
calculated the resulting luminosity of ionizing photons:
$Q_{H}=10^{38.13}$. In general it is not thought that the EUV flux of
M dwarfs is enhanced relative to the X-ray (Audard et al 2000), and
limits on the UV flux of this source have been obtained (Christian et
al 1999). Thus, the distance of this source would be 10 times closer
with an ionizing output six orders of magnitude lower than one would
require given the considerations in Sec. 3.4.

Second, the proper motion measured for these stars yields a
transverse velocity of $40~ {\rm km~s^{-1}}$ (Odenkirchen et al. 1997),
with a position angle of the proper motion is $248^{\circ}$, while
the filament has a position angle of $\sim 184^{\circ}$. The proper
motions of these two stars and many others near the southern end
of the filament are shown in Figure 4. 
(Odenkirchen et al. measured 365 stars' proper motions because they
were near M81, which happens to be near the southern end of our filament.)
We describe
in the next paragraphs that these two position angles cannot be
reconciled given what we know about the relative motions of the Sun,
this binary system, and the local interstellar gas.

Imagine that the ISM is stationary with respect to the local standard
of rest (LSR).  If the Sun also was stationary with respect to the
LSR, then the proper motion of the WD that made the FST would be
aligned with the FST.  Because the Sun moves with respect to the LSR,
we should transform the WD's heliocentric radial velocity and proper
motion to the LSR frame before comparing the WD's path to the
H$\alpha$~filament.  Two determinations of the solar motion,
or equivalently two definitions of the LSR, are that the Sun
is moving at +20 km~s$^{-1}$ towards $(l,b) = (56,+23)$ (Delhaye 1965)
or that the Sun is moving at +13 km~s$^{-1}$ towards $(l,b) =
(28,+32)$ (Dehnen \& Binney 1998).

In actuality, the solar motion, whatever its value and
direction, is not a sufficient correction to the WD's motion, because
an interstellar wind blows with respect to the LSR.  That is, the LSR
is defined in terms of the motions of stars, and interstellar gases
move with respect to that LSR.

As measured from a reference frame at rest with respect to the local
wind, the proper motion of the WD will be aligned with the FST.  (This
assumes that the wind near the Sun is identical to the wind at the
WD's location.)  Two determinations of the local ISM's wind are that a
wind is coming toward the heliocentric reference frame at 28
km~s$^{-1}$ from $(l,b) = (25,+10)$ (Crutcher 1982), or 23 km~s$^{-1}$
from $(l,b) = (355,-21)$ (Holzer 1989)
Crutcher analyzed optical absorption lines
and 21-cm emission lines over distances of approximately 100 pc.
Holzer reviews results based upon backscattered light from the
heliopause, at a distance of approximately 100 AU.

We have transformed the heliocentric velocity vectors of several candidate
stars as described above in order to predict the locations of those
stars in the past few hundred thousand years. Having performed this
transformation for each of the four possible
reference frames described above,
we cannot match the filament's location and position angle with
either of the two candidates chosen based upon their location in line
with and south of the filament: the double system LHS 2176/2178; or
the brightest star in Figure 1 that lies along the filament, SAO 14966.

We are left, therefore, with one candidate ionizing source, but no
convincing optical counterpart.  We note, however, that a white dwarf
with an ionizing luminosity of $Q_{H} \leq 10^{44}~{\rm s^{-1}}$ would have
an absolute V magnitude of approximately 10, so that if the 
distance, $d \geq 100$ pc, the blue-colored hot star would have an
apparent visual magnitude  $\geq 15$. Such an object might have been overlooked.

\subsection{Depolarization of Background Radio Emission}

Recent observations of the polarization of the ISM between 325 MHz and 5 GHz
reveal a class of relatively straight, filamentary
structures seen in Faraday depolarization of synchrotron emission from
the Galactic halo
(c.f., Haverkorn, Katgert, \& de Bruyn 2000; Wieringa et al. 1993). This
presumably occurs when the Faraday rotation across the beam is large
enough that the net polarization drops. We are not aware of any such
observations specifically of our filament. Observations of Haverkorn et al.
(2000) have an angular resolution of 4\arcmin, and their filaments (the Dutch
authors call them ``canals'') appear essentially unresolved.
The angular lengths of those filaments are similar to our object. For reasonable
interstellar parameters, the filament described in this paper could
produce a Faraday  depolarization. The angle of rotation (in radians) is
given by $\theta_{rot}=0.81 \lambda_{m}^{2} \int B_{|| \mu G} n_{e}
dl_{pc}$, where $\lambda_{m}$ is the observed wavelength in meters, and
$n_{e}$ is the electron density in ${\rm cm^{-3}}$ (Spitzer 1978).
If we make the
assumption that the density and magnetic field are uniform in the
filament, 
then the rotation of the angle of polarization can be written as

\begin{equation}
\theta_{rot}=1.9~B_{|| \mu G}~n_{e}^{-1}~I_{H\alpha}~T_{4}^{0.92}
\left[ \frac{\nu}{325~{\rm MHz}} \right]^{-2} {\rm radians}
\end{equation}

In this equation, $I_{H\alpha}$ is measured in rayleighs. At 325 MHz, for
a brightness of 0.5 rayleigh, an assumed magnetic field strength of 3 $\mu
G$, and a temperature of 10,000 K, a rotation of 1 radian would
occur for a density of $\sim$ 3 ${\rm cm^{-3}}$. If a connection
between filaments seen in emission and polarization is established, a
comparison of the two observations can constrain the ratio of magnetic
field strength to gas density.

\section{Future Observational Tests} \label{tests}

Although the extreme linearity of the feature seems evidence
for a point-source origin for this filament; it remains possible
that is just an ordinary nebular filamentation that is
unusually straight. Here, we discuss possible tests by which the
nature of this filament might be ascertained.

There are several generic features that we expect to be associated
with a FST that would not be present for a standard nebular
filament. First, as the hydrogen recombines, there should be an
intensity gradient in $H \alpha$ along the filament. Second, since
different ions recombine at different rates, there should be a
significant variation in line ratios along the filament. Third, since
the gas has been impulsively heated by photoionization from the passing
source, there should be a decrease in gas temperature along the
filament. Fourth, the heating should also produce an thermally
overpressured region that will expand along the filament (depending
upon the ambient pressure structure). Finally, random motions in the
surrounding medium should produce irregularities and distortion of the
trail over time. All of these feature are amenable to observational
tests. However, it may be difficult to sort out some of these
effects. Complicating the issues is that fact that the luminosity and temperature of a WD can vary considerably along the trail. Here we present some of the more fundamental considerations. A detailed model is deferred to future work.

{\it Intensity gradients:} Probably the most compelling test would be
to search for ionization and intensity gradients along the filament,
using narrow band imaging or spectroscopy.
If the filament were unusually straight by
chance alone, we would expect that the intensity and ionization level
should be roughly constant along the filament. On the other hand, if
the filament were an example of a shocked or photoionized trail, one
would expect to see a variation in the intensity and ionization level,
and perhaps a characteristic velocity pattern as well.

If we define $\theta=0$ (measured in
degrees) as the current position of the ionizing source, then the electron
density is given by 

\begin{equation}
n_{e}(\theta)=(n_{e}^{-1}(0) + 0.14
T_{4}^{-0.8}v_{100}^{-1}d_{100}p\theta )^{-1}~.
\end{equation}

This will result in a gradient in intensity along the length of the
FST.  In the actual interstellar medium, however, the absolute
intensity along the FST will be affected by inhomegenities in the
ambient density.  

{\it Ionization gradients:} A method that would be insensitive to
inhomogeneities would be to measure the ratio of intensities of
different lines, such as the forbidden optical transitions of [S II]
and [N II], as a function of position.  If the ionization is
impulsive, and the gas is recombining behind some source of shock
ionization or photoionization, there will be a variation in the line
ratios with position.  The
recombination time for N II is 58\% of that of H II, while S II is
250\% longer than H II (Raymond \& Smith
1977; Raymond \& Cox 1985) . The Case A recombination time of H II is
$t(HII)_{rec}=0.077 \times 10^{6} n_{e}^{-1}$ years. Figure 5 shows the
intensities of the [S II], [N II], and H$\alpha$ transitions as
functions of position along the FST .  The intensities have all been
normalized to unity at $\theta=0$, and the initial state of the gas is
assumed to be fully ionized, with nitrogen and sulfur all in the first
ionization stage. This assumes the gas temperature is constant at $T=10^{4}$ K. 

In addition to providing us with evidence that there is or is not an
ionization gradient, additional observations may yield valuable
information on the ionization mechanism. For example, the Guitar
nebula emits a ``nonradiative'' shock spectrum of pure Balmer lines
and no metal lines, so detecting S II from our object would prove it
is not the same type of object as the Guitar nebula. If the spectrum
looks like typical ISM spectra, e.g. with a S/H line intensity ratio
of 0.4 (Reynolds 1985) independent of position, then it is probably
just another faint filament in the interstellar medium that happens to
be at moderate galactic latitude and is unusually thin, straight, and
long.  The intensity ratio of S~II~6716~/~H$\alpha$~6563 is 0.1 in
\stromgren~spheres, and 0.4 in the diffuse ISM (Reynolds 1985).  The
intensity ratio of N~II~6584~/~H$\alpha$~6563 is similar to
S~II~6716~/~H$\alpha$~6563 in the diffuse ISM (Haffner, Reynolds, \&
Tufte 1999).
Scaling from the 0.5 R H$\alpha$~line, the expected intensity of the S
II 6716 line will be 0.05 to 0.2 R, depending on the characteristics
of the ionization source.

{\it Temperature gradients:} The cooling rate for solar metallicity
plasma at $T=10^{4}$ K is 
$\Lambda=n_{e}~n_{H}10^{-24}~{\rm~ergs~cm^{-3}s^{-1}}$ (Osterbrock 1989).
Assuming a constant pressure, the
cooling time is $t_{cool}=T/(dT/dt)=0.25 \times 10^{6} n_{e}^{-1} $
years. Since this time scale is comparable to the timescale for
recombination,  collisionally excited optical line emission may be suppressed
compared
to the considerations presented above, which considered only the ionization
balance at a constant temperature. A potential further complication is that
adiabatic expansion as well as radiative cooling may contribute to the energy
loss.

{\it Nebular expansion:} Given that the gas is heated by a source that
is (probably) moving supersonically with respect to the neutral
interstellar medium, the heating will be instantaneous compared to the
subsequent cooling and dynamical readjustment. Clearly, the filament
we have detected has this morphology, being wider at one end than the
other. If we think of the Y-shape as a V-shape, then we can define an
opening angle, $\beta$, of the two line segments of the V.
The observed value of $\beta_{obs}~\cong~$5\arcmin/2.5\arcdeg~=~0.033 radians.
If we assume that the filament lies in
the plane of the sky and expands transversely at approximately the sound
speed, $c_{s}$ of the heated ($T=10^{4}$ K) gas, then 
$\beta~\cong~2~c_{s}/v_{source}~\cong~0.25~T_{4}^{1/2}~v_{100}^{-1}$~radians.
This is rather large compared to the observed opening angle,
unless the source velocity were extremely high.  The
assumption that the gas expands at the sound speed may be somewhat extreme,
because other sources of interstellar pressure may dominate
over the thermal pressure. Evidence
for position dependent expansion velocities or velocity gradients would
provide important constraints for more detailed ionization and
dynamical modeling that are beyond the scope of this paper.

\section{Summary}

We report on the discovery of a narrow, 2.5\arcdeg~long, ionized filament
in the ISM
that looks much like the contrail behind an aircraft. While its origin
may be no different than any other such filament in the ISM, and
its straightness and its narrowness are simply a fluke, another possibility
is that it is the trail left by a white dwarf or some other low-luminosity
source moving quickly through a neutral medium and photoionizing it
as it passes. We explore the implications of such a model, which may be
called a ``\stromgren~cylinder'' or a ``Fossil Stromgren Trail.''
We parameterize its length, width, and brightness in terms of the
ionizing source's luminosity and velocity, and the ambient gas density.
Two other models, namely a stellar jet and a shock-ionized trail,
are noted but found to be unlikely explanations because this filament's
properties are very different than other examples of those types of
object.

Whatever the cause of the filament, it
deserves further study based upon its unique appearance. Its large angular
size suggests that
it is nearby, and therefore it should be possible to identify what
has created it. For example, if the culprit star can be identified,
then this filament may be
used to constrain interesting parameters of the local interstellar medium:
in particular,
the density of the ambient gas and its motion with respect to the LSR.
A particularly interesting question is what determines the length of the
trail. Is it dominated by local microphysics, i.e., the hydrogen
recombination time, as we have assumed, or is it dissipation due to
shear or turbulent motions in the ISM? The longevity of the trail
therefore could be used to constrain the velocity dispersion in the local ISM.
If filaments like this one can be shown to be ``streak lines''\footnote{In
hydrodynamics, a streakline is a line traced out by a dye continuously
injected into the flow.}
in the flow of interstellar gas, then collectively and individually they would 
be a significant new probe of the dynamics of the interstellar
medium anywhere that they are found.

\acknowledgments

We thank the staff of the Mt. Laguna Observatory:
Bob Leach built the CCD electronics used in the camera;
Jay Grover mounted the camera on the side of the 1-m telescope. 
We would also like to thank Greg Madsen, Ron Reynolds and Matt Haffner
for obtaining an $H \alpha$ image (not shown here) of this filament. 
M. Odenkirchen kindly provided the proper motions of stars near M81
in machine-readable form.
The camera, filter wheel, and filters were paid for in part
by an Alfred P. Sloan Research Fellowship to P. R. M., 
who is supported in part by a Cottrell Scholarship from the Research
Corporation and by the NSF CAREER award AST-9874670. R. A. B
would like to acknowledge support from NASA Astrophysical Theory Grant
NAG5-8417. We also gratefully acknowledge the use of Skyview, maintained by the Laboratory for High Energy Astrophysics at NASA/GSFC  and SIMBAD, operated by the Centre de Donnees Astronomiques de Strasbourg in this research.

\clearpage

\begin{figure}
\epsscale{0.5}
\plotone{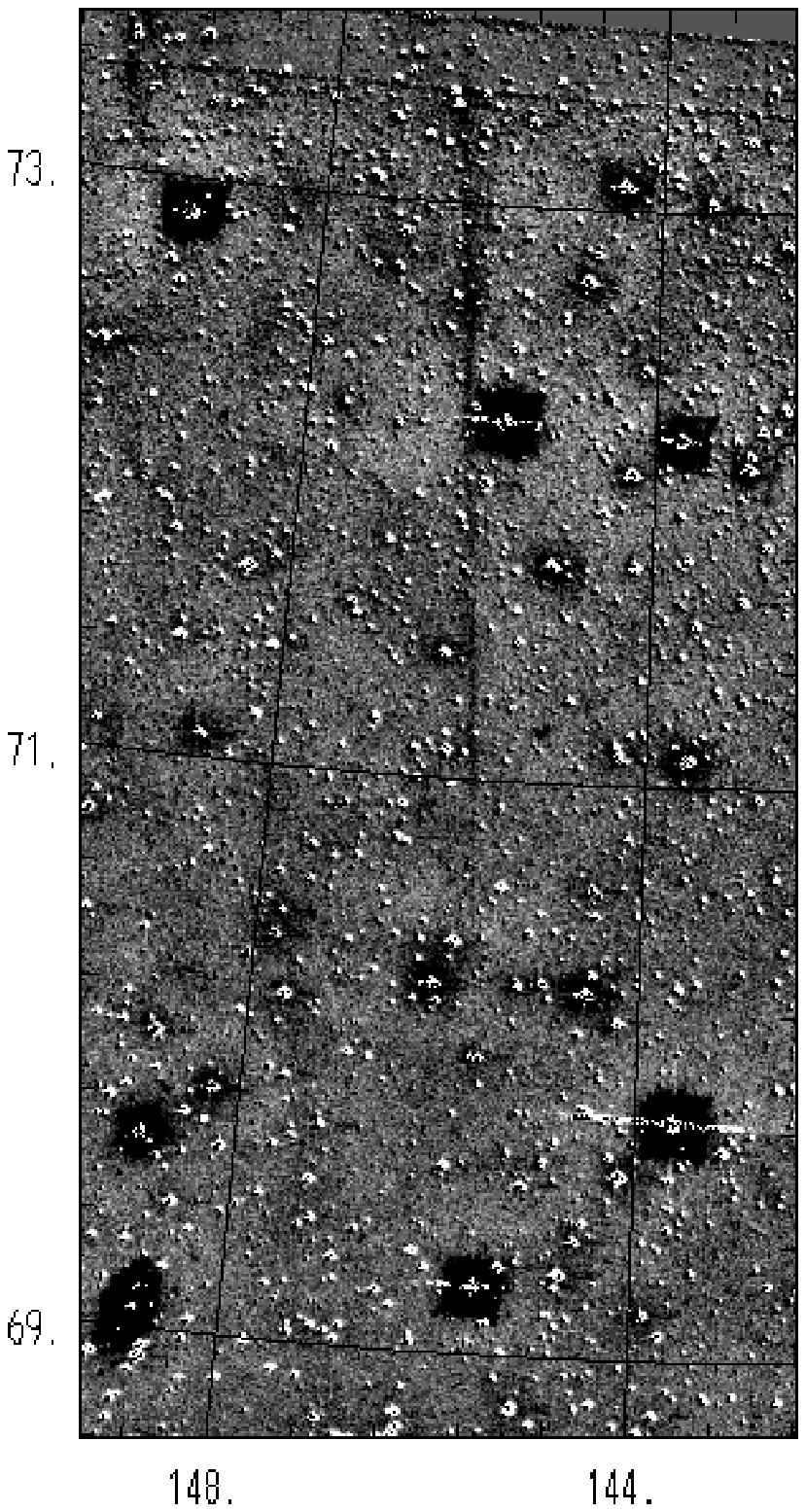}
\caption{This continuum-subtracted H$\alpha$ image shows
the long, linear, ionized filament that is the focus of this paper;
it extends from the center to the top of the image. The image is
a negative: bright objects appear black. The filament's H$\alpha$~surface
brightness is ${\sim}0.5$ rayleighs above background. The J2000 right
ascension and declination, ($\alpha$,$\delta$), are both in degrees.
One pixel corresponds to 0.265\arcmin and the angular resolution is
2.0 pixels (FWHM). The r.m.s. noise in each pixel is measured to
be 0.57 rayleighs in ``blank'' regions of sky.
One of the stars discussed in the text, SAO 14966, is the saturated
black blob at ($\alpha$,$\delta$) = (145.5\arcdeg,69.2\arcdeg).
The nearly horizontal line at $\delta = 73.3$\arcdeg~is an artifact of
the edge of one of the three images that were combined to make this image.
The filament appears in each of the three 1800-second exposures,
which were dithered
on the sky by approximately 10\arcmin with respect to each other.
The galaxy M81 is located at (148.8\arcdeg,69.1\arcdeg).
\label{fig1}}
\end{figure}

\begin{figure}
\plotone{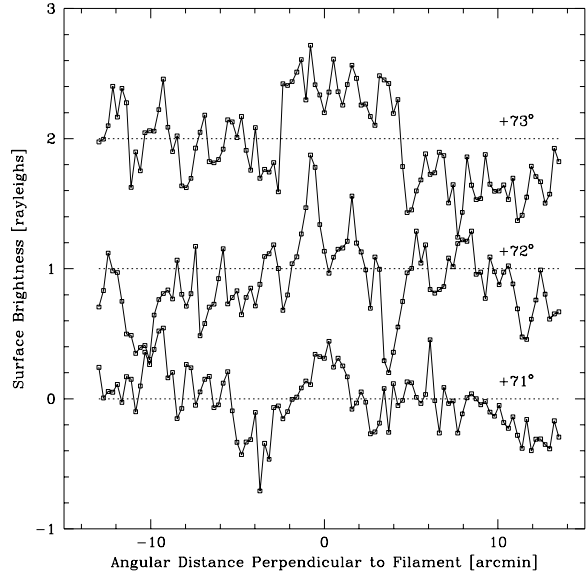}
\caption{These three cross sections of the filament in Figure 1, taken
perpendicular to its long axis at declinations of +71\arcdeg,
+72\arcdeg, and +73\arcdeg, show the filament's surface brightness in
rayleighs. The lines have been shifted vertically for clarity; the dotted
lines show the zero levels. The filament nominally is in the center of each
plot; at $\delta = +73$\arcdeg~its peak is displaced to the left (East)
by 1\arcmin.
Before extracting the cross sections, we median filtered the
image in Figure 1 with a kernel of 1x21 pixels, i.e. with a 21-pixel-long
rectangle oriented parallel to the filament's long axis. This process
removes stars without affecting the filament much. The widening of the
filament at $\delta = +73$\arcdeg~is apparent in the top slice; its
detailed structure is shown in the next Figure.
\label{fig2}}
\end{figure}

\begin{figure}
\plottwo{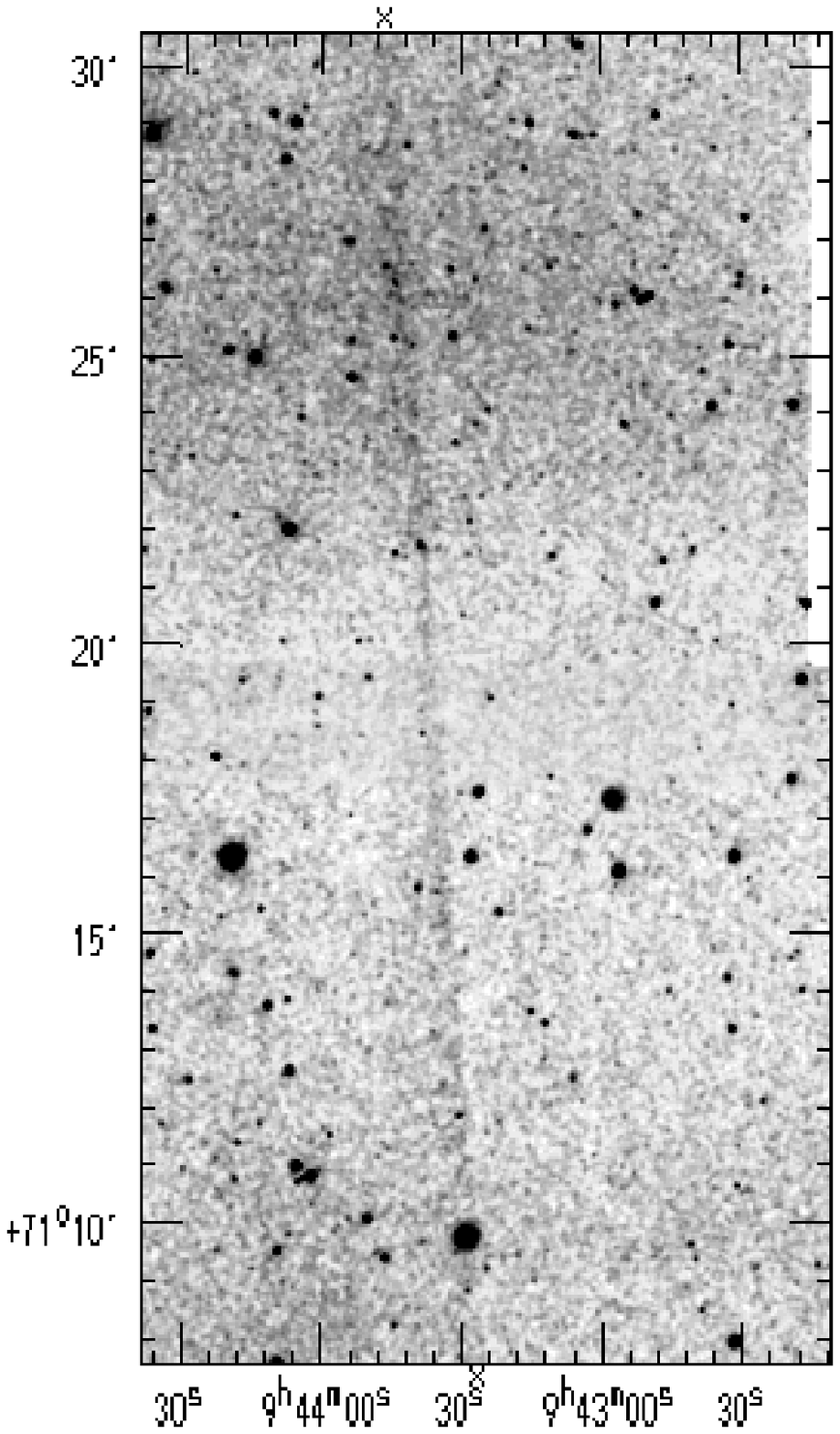}{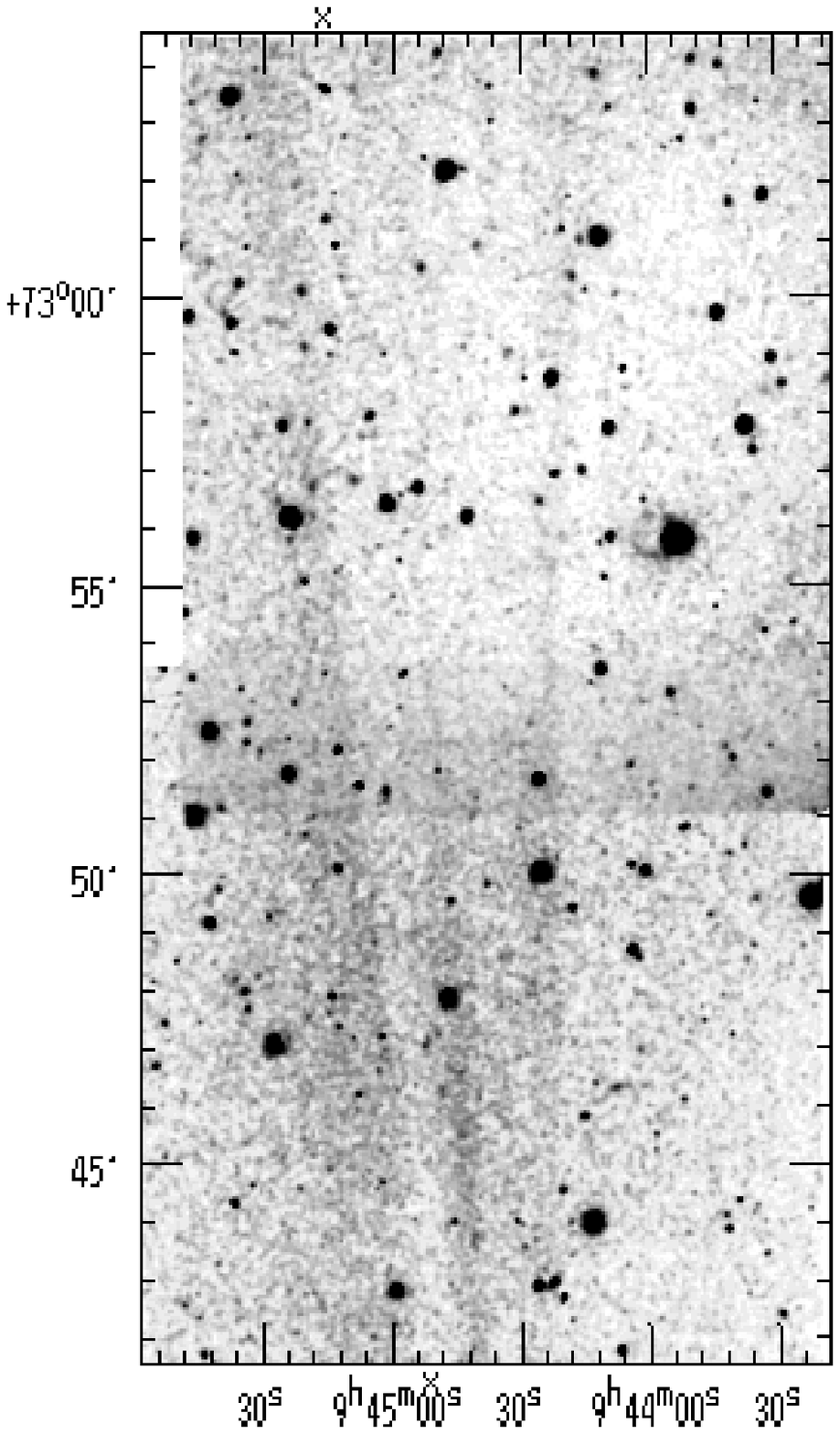}
\caption{These H$\alpha$ images shows the southern (left) and northern (right)
portions of the filament at higher angular resolution (FWHM = 0.1$^{\prime}$)
than Figure 1 (FWHM = 0.53$^{\prime}$). The filament is less than
0.5$^{\prime}$ wide in the left image, and flares into three components
separated at most by 5$^{\prime}$ in the right image.
Each of the two rectangular images is a mosaic of two overlapping square images,
each of which was a 1800 second exposure made with the Mt. Laguna 1-m diameter
telescope.
The filament is marked by two Xs on the border of the left image;
the line defined by those two Xs, if extended
as a great circle on the sky, would cross the right image as indicated by
the pair of Xs on that image. 
To compare these images, in which North is up and East is to the left, 
to those in Figure 1, note that the image in Figure 1 is rotated to orient
the filament vertically.
\label{fig3}}
\end{figure}

\begin{figure}
\epsscale{1.0}
\plotone{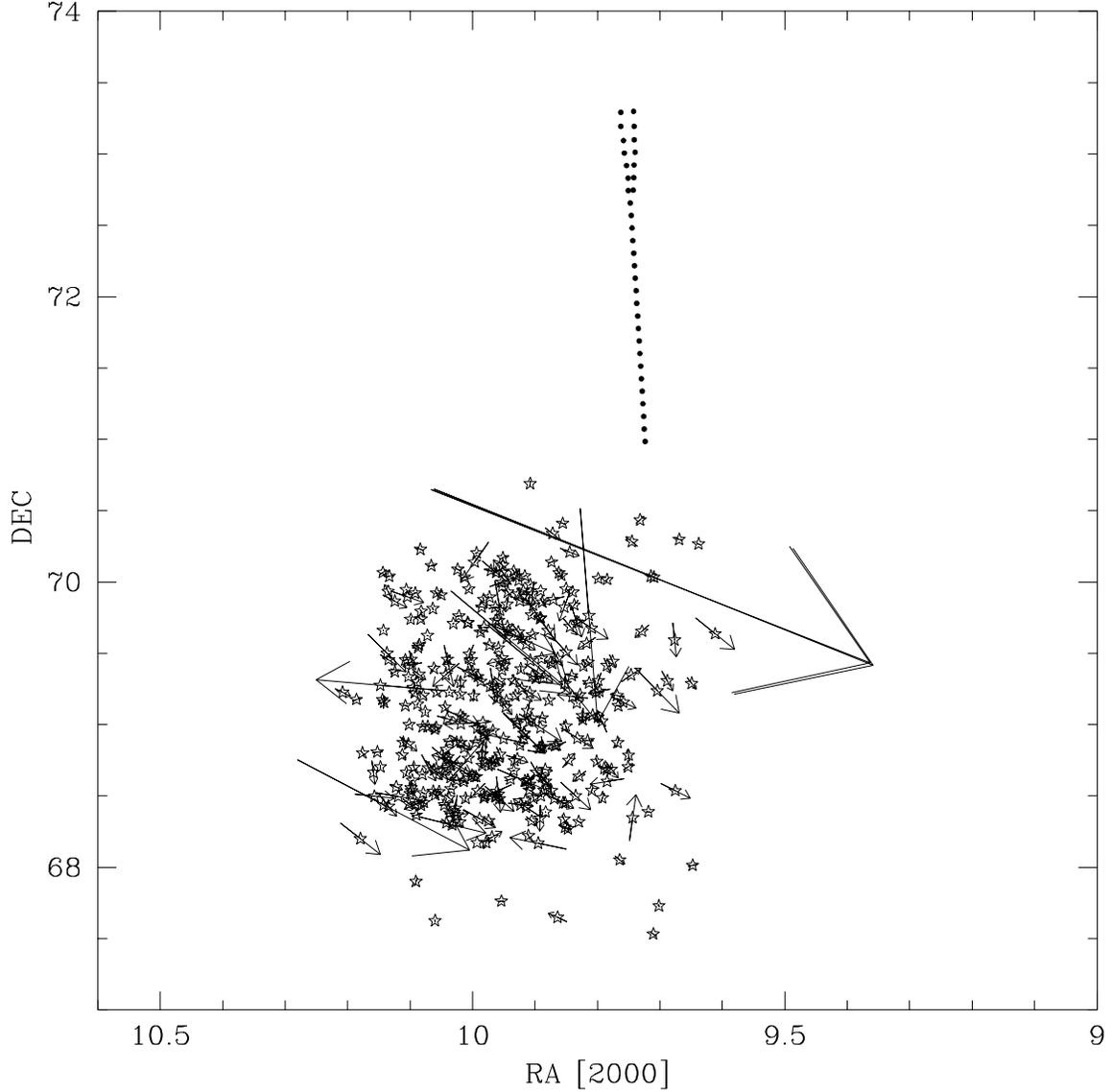}
\caption{The locations and proper motion of stars measured by
Odenkirchen {\it et al.} (1997) are plotted below the Y-shaped
filament discussed in this paper.
The largest proper motion of these 365 stars is that of the binary
system, LHS 2176/LHS 2178, with an angular rate of
0.72$^{\prime \prime}$/year. Located at 9$^h$.71, $+70.0$\arcdeg,
that binary is almost exactly in line with the filament, and it may be
associated with an extreme-UV and X-ray source (see text), but its
proper motion is at a position angle of 248\arcdeg, whereas the position
angle of the filament is 184\arcdeg.
\label{fig4}}
\end{figure}

\begin{figure}
\plotfiddle{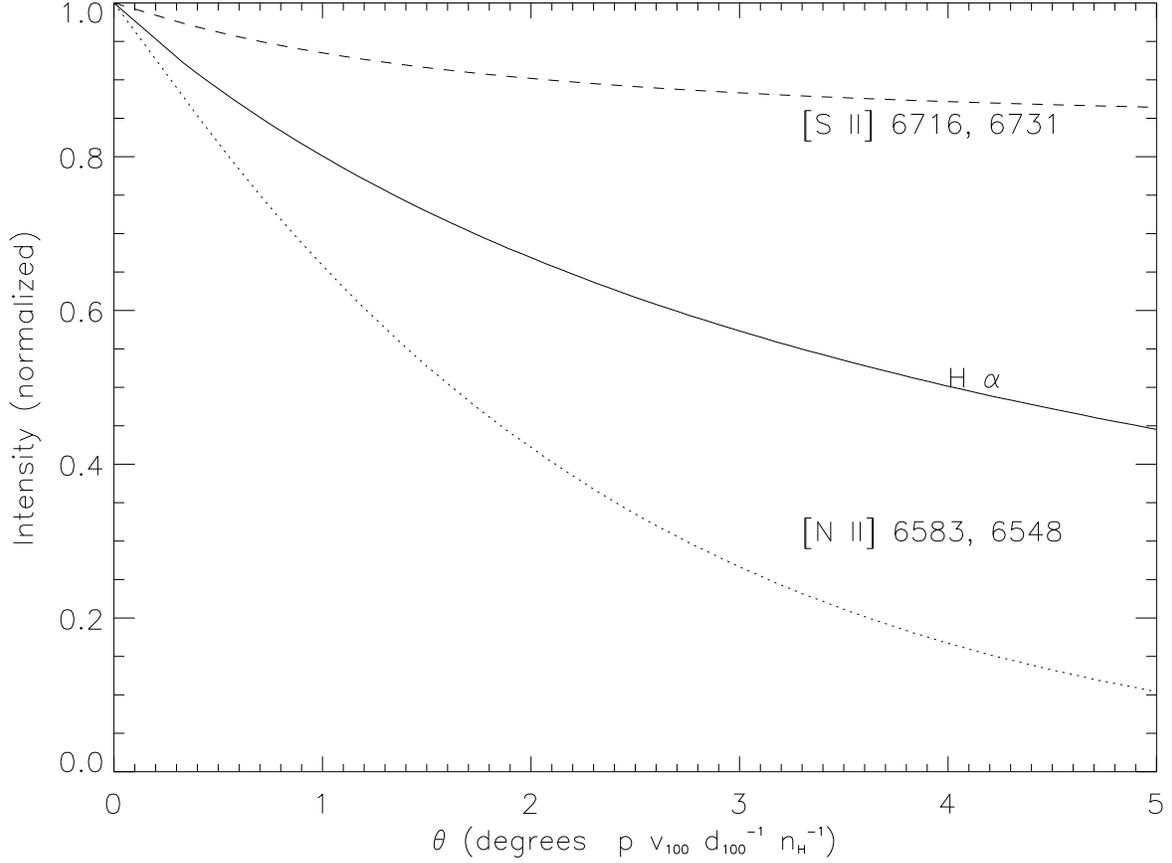}{340pt}{90}{70}{70}{270pt}{0pt}
\caption{ Variation of normalized optical emission line intensity for a
``Fossil \stromgren~Trail'' as a function of angle, $\theta$. The value of
the angle is normalized to case with source distance of 100 pc, source
velocity of 100 ~${\rm km~s^{-1}}$ and projection factor, $p$, in a
medium with particle density $n_{H}=1~{\rm cm^{-3}}$ and gas temperature
$T=10^{4}$ K. This assumes that hydrogen is initially fully ionized,
while sulfur and nitrogen are completely in the first ionization stage,
and that the temperature is held constant at $T=10^{4}$ K. 
\label{fig5}}
\end{figure}

\end{document}